\documentclass{SCGE}
\usepackage{multicol}
\begin{document}

\begin{picture}(0,0){\rm
\put(0,-39){\makebox[160truemm][l]{\bf {\sanhao\raisebox{2pt}{.}}
Research Paper  {\sanhao\raisebox{1.5pt}{.}}}}}
\end{picture}

\def\bm{\boldsymbol}

\def\dl{\displaystyle}
\def\du{\end{document}}

\Year{2012} %
\Month{May}
\Vol{55} %
\No{5} %
\BeginPage{872} %
\EndPage{879} %
\AuthorMark{{\rm DONG Yao,} et al.}
\DOI{DOI: 10.1007/s11433-012-4707-8} 

\title{Tidal evolution of exo-planetary systems: WASP-50, GJ 1214 and CoRoT-7}

\author[1,2]{DONG Yao}{}
\author[1*]{JI Jianghui}{}

\address[{\rm1}]{Purple Mountain Observatory, Chinese Academy of
Sciences, Nanjing {\rm 210008}, China;}
\address[{\rm2}]{Graduate University of Chinese Academy of Sciences, Beijing
100049, China}

\maketitle \vspace{-3.5mm}{\footnotesize\begin{center} Received ; accepted 
\end{center}}\vspace*{-5mm}

\begin{center}
\rule{16.5cm}{0.4pt}
\parbox{16.5cm}
{\begin{abstract} We perform numerical simulations to investigate
tidal evolution of two single-planet systems, that is, WASP-50 and
GJ 1214 and a two-planet system CoRoT-7. The results of orbital
evolution show that tidal decay and circularization may play a
significant role in shaping their final orbits, which is related to
the initial orbital data in the simulations. For GJ 1214 system,
different cases of initial eccentricity are also considered as only
an upper limit of its eccentricity (0.27) is shown, and the outcome
suggests a possible maximum initial eccentricity (0.4) in the
adopted dynamical model. Moreover, additional runs with alternative
values of dissipation factor $Q^\prime_1$ are carried out to explore
tidal evolution for GJ 1214b, and these results further indicate
that the real $Q^\prime_1$ of GJ 1214b may be much larger than its
typical value, which may reasonably suggest that GJ 1214b bears a
present-day larger eccentricity, undergoing tidal circularization at
a slow rate. For the CoRoT-7 system, tidal forces make two planets
migrating towards their host star as well as producing tidal
circularization, and in this process tidal effects and mutual
gravitational interactions are coupled with each other. Various
scenarios of the initial eccentricity of the outer planet have also
been done to investigate final planetary configuration. Tidal decay
arising from stellar tides may still work for each system as the
eccentricity decreases to zero, and this is in association with the
remaining lifetime of
each planet used to predict its future.
\end{abstract}}
\end{center}\vspace*{-0.6cm}
\begin{center}
\parbox{16.5cm}{\bf\jiuhao extrasolar planets, tidal decay, planetary formation, numerical simulations
}

\end{center}

\begin{center}
\parbox{16.5cm}{\PACS{\hspace*{-2mm}\rm 96.15.Wx, 96.15.Bc, 96.15.De}

\rule{16.5cm}{0.4pt}}\end{center}



\wuhao\vspace*{1.5mm}
\begin{multicols}{2}
\renewcommand{\baselinestretch}{1.08} \baselineskip 12.2pt\parindent=10.8pt

\no 

\section{Introduction}
There are 758 extrasolar planets detected up to 6 February 2012, a
large portion of which are Jupiter-like planets, as well as a
population of super-earths with masses ranging from 1 to 10 Earth's
mass ($M_\oplus$) [1]. Figure 1 shows the distribution of semi-major
axes $a$ and eccentricities $e$ of all observed extrasolar planets
by Doppler radial velocity. The eccentricities of the planets tend
to be close to zero for very close-in orbits ($a$ $<$ 0.2), however,
the eccentricities of planets are distributed uniformly from 0 to
0.9 when $a$ $>$ 0.2, which may be explained as a result of tidal
decay and circularization in the exo-planetary systems. As well
known, tidal dissipation is quite common in the planetary systems.
For example, in our solar system, tidal effects existing in the
Moon-Earth system may make the Moon gradually move away from the
Earth,

\noindent\rule{2.5cm}{0.4pt}\\[0.1mm]{\qihao *Corresponding author (email:
jijh@pmo.ac.cn)\\} and at the same time the rotation of Earth becomes a
bit slower. In addition, tidal heating of Io (one of Galilean moons
of Jupiter) changes the satellite's interior. Moreover, some planets
and most satellites have been in synchronous rotation due to
tidal dissipation [2].

In exo-planetary systems, the distribution of semi-major axes has a
cut-off (0.01 AU) beyond which no planet is unveiled, because very
close-in planets have been destroyed by tidal effects. The
distribution of the planetary ages and the semi-major axes further
shows that less younger planets have a much smaller semi-major axes
because most of them have been engulfed by the host star due to the
stronger tides [3].

Herein Jupiter-like planets (super-earths) with semi-major axes $a
<$ 0.1 AU, are called hot Jupiter-like planets (hot super-earths)
[4]. However, it is difficult to elucidate how these planets
have formed in such close-in orbits as observed according to current
planetary formation models [5-9]. However, a possible dynamical
evolution picture may be described as follows: originally, the
planets form in the distant proto-planetary gaseous disk (several
AU), then move into closer orbits due to the migration mechanisms
such as disk migration [10,11], the secular stellar perturbation or
Kozai mechanism [12] and planetary scattering (planets will be
usually excited to high eccentricities) [13]. Once they move to a
much closer orbit especially within 0.1 AU, tidal effects arising
from its host star then plays a vital role in the secular evolution
of the planets, to shape their final orbits [6]. Finally, it may
come into being the observed planetary configuration as tidal decay
and circularization [14].

For multiple-planet systems, consisting of at least two planets, it
is believed that gravitational interaction between planets is
coupled with tidal process, so that this may cause the orbital
evolution to be more complex than that of single-planet systems. The
previous works are not well studied for such systems [15,16].
Therefore, we not only investigate tidal evolution of two
single-planet systems (WASP-50 and GJ 1214), but also one two-planet
system (CoRoT-7). When a planet has a nonzero eccentricity,
planetary tide (arising from the central stellar acting on a planet)
will decrease its semi-major axis and eccentricity until the
eccentricity drops to zero, including current observed orbits in
this process. The stellar tide (arising from the planet acting on
the central star) continues to cause the planet's orbit to decay
even after the eccentricity decreases to zero, which is associated
with the remaining lifetime of each planet to predict its future.

\begin{center}
 \centerline{\psfig{figure=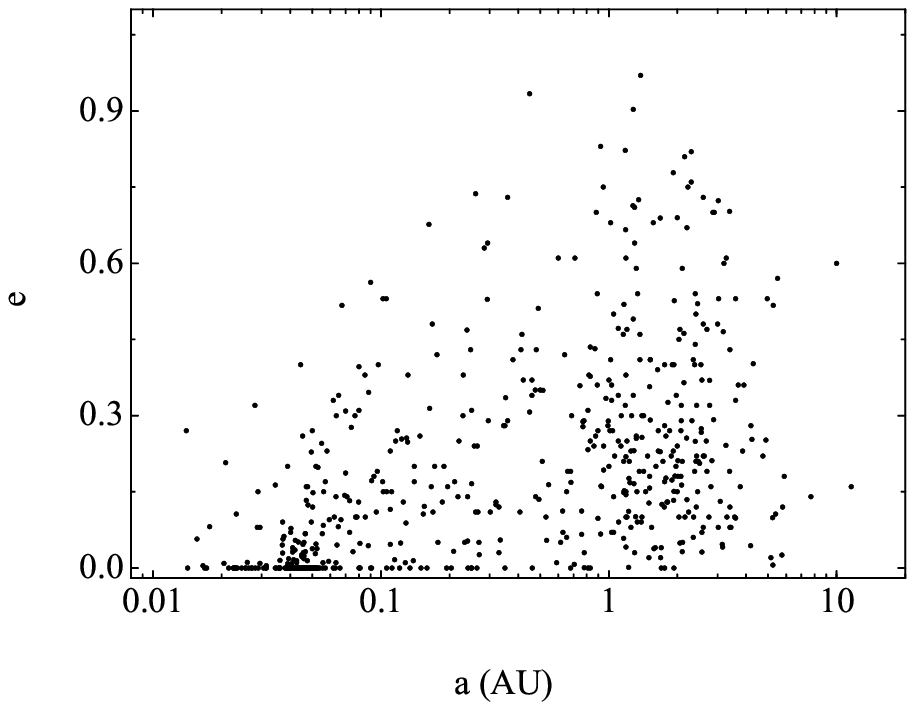,width=3.2in}}\end{center}\vspace*{-8.5mm}
 {\footnotesize {\bf Figure 1}
 \quad The distribution of the eccentricities $e$ and semi-major axes $a$ [1] for extrasolar planets.}

\section{Tidal Theory}

According to the classical tidal theory, the mean variations of the
semi-major axis and eccentricity in a single-planet system are given
by [16,17]

\begin{equation}
 < \frac{da}{dt} > = -\frac{4}{3}na^{-4}\hat{s}[(1+23e^2)+7e^2D]
 \end{equation}
 and
 \begin{equation}
 < \frac{de}{dt} > = -\frac{2}{3}nea^{-5}\hat{s}[9+7D]
 \end{equation}
where $D\equiv$ $\hat{p}/2\hat{s}$, and

 \begin{equation}
\hat{s} \equiv \frac{9}{4}\frac{k_0}{Q_0}\frac{m_1}{m_0}R_0^5,
\quad\hat{p} \equiv \frac{9}{2}\frac{k_1}{Q_1}\frac{m_0}{m_1}R_1^5,
 \end{equation}
represent the stellar and planetary tides respectively. In the above
equations, $m_i$, $R_i$ ($i$ = 0, 1) are the mass and radius (in
this paper, the subscripts 0 and 1 mean stellar and planetary
parameters, for two-planet system, the subscripts 1 and 2 relate to the
inner and outer planet, respectively), $a$, $e$ and $n$ are
semi-major axis, eccentricity and mean orbital motion, and $k_i$,
$Q_i$ are Love number and tidal dissipation factor, where $k_i$ is
related to the inner tidal-effective rigidity, the distribution of
radial density and other parameters of a planet. $Q_i$ is a
characterization of the tidal strength, that is larger $Q_i$
corresponding to more weaker tidal dissipation, otherwise a more
stronger tidal interaction.

The above tidal theory shows that $a$ and $e$ are nonlinearly
coupled and cannot be treated independently as a simple function of
time. Furthermore, when the ratio of $m_1/m_0$ is smaller or $Q_0\gg
Q_1$, in these cases that stellar tide can be ignored. Hence, the
stellar tide is not taken into account in the planetary tidal
evolution until the planetary eccentricity drops to be zero in this
work. This indicates that the stellar tide predicts the remaining
lifetime of the planetary system.

\section{Numerical Simulations of Tidal Evolution}
\subsection{Dynamical Model}

The reference framework used here is centered at the star and the
motion of the planet is coplanar with respect to the reference
plane. General relativity (GR) is considered for all planets here
due to the planetary close-in orbits. Then the dynamical equations
for a two-planet system are given as the following form by adding GR
on the outer planet [16]:

\begin{eqnarray}
\ddot{\textbf{r}}_1&=&-\frac{G(m_0+m_1)}{r_1^3}\textbf{r}_1+Gm_2\Bigg{(}\frac{\textbf{r}_2-\textbf{r}_1}{|\textbf{r}_2-\textbf{r}_1|^3}-\frac{\textbf{r}_2}{r_2^3}\Bigg{)}\\
&&+\frac{(m_0+m_1)}{m_0m_1}(\textbf{\textbf{f}}_{\textrm{t}_1}+\textbf{f}_{\textrm{g}_1})+\frac{(\textbf{f}_{\textrm{t}_2}+\textbf{f}_{\textrm{g}_2})}{m_0},\nonumber\\
\ddot{\textbf{r}}_2&=&-\frac{G(m_0+m_2)}{r_2^3}\textbf{r}_2+Gm_1\Bigg{(}\frac{\textbf{r}_1-\textbf{r}_2}{|\textbf{r}_1-\textbf{r}_2|^3}-\frac{\textbf{r}_1}{r_1^3}\Bigg{)}\nonumber\\
&&+\frac{(m_0+m_2)}{m_0m_2}(\textbf{\textbf{f}}_{\textrm{t}_2}+\textbf{f}_{\textrm{g}_2})+\frac{(\textbf{f}_{\textrm{t}_1}+\textbf{f}_{\textrm{g}_1})}{m_0}
\end{eqnarray}
where $\textbf{f}_{\mathrm{g_1}}$ and $\textbf{f}_{\mathrm{g_2}}$
are forces acting on the inner and outer planets generated from GR
given by [16,18]

\begin{equation}
\textbf{\textbf{f}}_{\mathrm{g_1}} =
\frac{Gm_0m_1}{c^2r_1^3}\Big[\Big(4\frac{Gm_0}{r_1}-\textbf{v}_1^2\Big)\textbf{r}_1+4(\textbf{r}_1\cdot\textbf{v}_1)\textbf{v}_1\Big]
   \label{fg1}
\end{equation}

\begin{equation}
 \textbf{f}_{\mathrm{g_2}} = \frac{Gm_0m_2}{c^2r_2^3}\Big[\Big(4\frac{Gm_0}{r_2}-\textbf{v}_2^2\Big)\textbf{r}_2+4(\textbf{r}_2\cdot\textbf{v}_2)\textbf{v}_2\Big]
   \label{fg2}
\end{equation}
where $\textbf{v}_1$ = $\textbf{\.{r}}_1$ and $\textbf{v}_2$ =
$\textbf{\.{r}}_2$, are velocities of the inner and outer planets
respectively, and $c$ is the speed of light.

In the above equations, $\textbf{f}_{\mathrm{t_1}}$ and
$\textbf{f}_{\mathrm{t_2}}$ are tidal forces acting on the inner and
outer planet excited by the host star with the expressions of
~[16,19,20]

\begin{equation}
 \textbf{f}_{\mathrm{t_1}} = -\frac{9Gm_0^2R_1^5}{2Q_1^{'}n_1r_1^{10}}[2\textbf{r}_1(\textbf{r}_1\cdot\textbf{v}_1) + r_1^2(\textbf{r}_1\times\Omega_1+\textbf{v}_1)]
\end{equation}

\begin{equation}
 \textbf{f}_{\mathrm{t_2}} = -\frac{9Gm_0^2R_2^5}{2Q_2^{'}n_2r_2^{10}}[2\textbf{r}_2(\textbf{r}_2\cdot\textbf{v}_2) + r_2^2(\textbf{r}_2\times\Omega_2+\textbf{v}_2)]
\end{equation}
where $G$ is gravitational constant, $\Omega_i$ is the spin
velocity, and $Q^\prime_i$ denotes a modified tidal dissipation
factor defined as $Q^\prime_i \equiv 3Q_i/2k_i$, which is related to
the lag time of the deformation due to tidal interaction. In
general, $Q^\prime_i$ = 100 is adopted for Earth-like planets, e.g.
CoRoT-7b, but $10^5-10^6$ for Jupiter-like planets, for example, WASP-50b.

In this work, we adopt a modified MERCURY6 package to simulate the
orbital evolution by adding GR and tidal effect in the original
codes [23], which is a general-purpose software package for N-body
integrations and is designed to explore the dynamical evolution of
objects moving in the gravitational field of a massive central body,
as well to consider non-gravitational (e.g., cometary jet) process.
Bulirsch-Stoer algorithm is utilized in the numerical simulations,
by nature being slow but more accurate in the runs. The time
step adopted here is 1/50--1/100 times the period of the innermost
planet, with an accuracy of 10$^{-12}$--10$^{-16}$.

\subsection{WASP-50 System}

WASP-50b is a hot Jupiter discovered by transit photometry in 2011
residing at a close-in orbit (0.0295 AU) around a G9 dwarf ( $v$ =
11.6, 0.892 $M_{\odot}$, 0.843 $R_{\odot}$) (Table 1) [24]. The
current stellar activity and rotational period indicate that the age
of the host star is about 0.8 Gyr, which is discrepant with the
estimated age about 7 Gyr from stellar evolution. It may be
relevant to tidal dissipation [24]. WASP-50b might be in the final
process of undergoing tidal decay and circularization according to
its current extremely tiny eccentricity (0.009). Considering the
angular momentum of the system invariable throughout tidal
evolution, and the initial region before migration due to tidal
decay being more distant than the current location (the same
situation for other systems), then the initial orbital elements in
the numerical simulations are assumed as
$a_{1~\mathrm{\mathrm{ini}}}$ = 0.0305 AU, $e_{1~\mathrm{ini}}$ =
0.18. Herein a typical tidal dissipation factor $Q^\prime_1$ is
taken as $10^5$.

\vspace{4mm}\noindent {\small Table 1\quad Orbital and physical parameters of WASP-50 system [24]}\\
\vspace{-8mm} {\footnotesize
\begin{center}
\begin{tabular}{lccccc}
\hline
Body     &  Mass                & Radius             & Semi-major  axis  &  Eccentricity \\
\hline
 WASP-50 & 0.892   $M_{\odot}$   &   0.843 $R_{\odot}$ &$-$  &    $-$  \\
 WASP-50b & 1.468   $M_{J}$  &  1.153 $R_{J}$ & 0.0295 AU &  $<$0.009  \\
 \hline
\end{tabular}
\end{center}}\vspace{4mm}

Figure 2 shows that the variations of the semi-major axis and
eccentricity for WASP-50b. As expected, the semi-major axis and
eccentricity decrease over secular evolution, and finally reach
present location at 0.0295 AU in a nearly-circular orbit within
$\sim$ 50 Myr.

\begin{center}
 \centerline{\psfig{figure=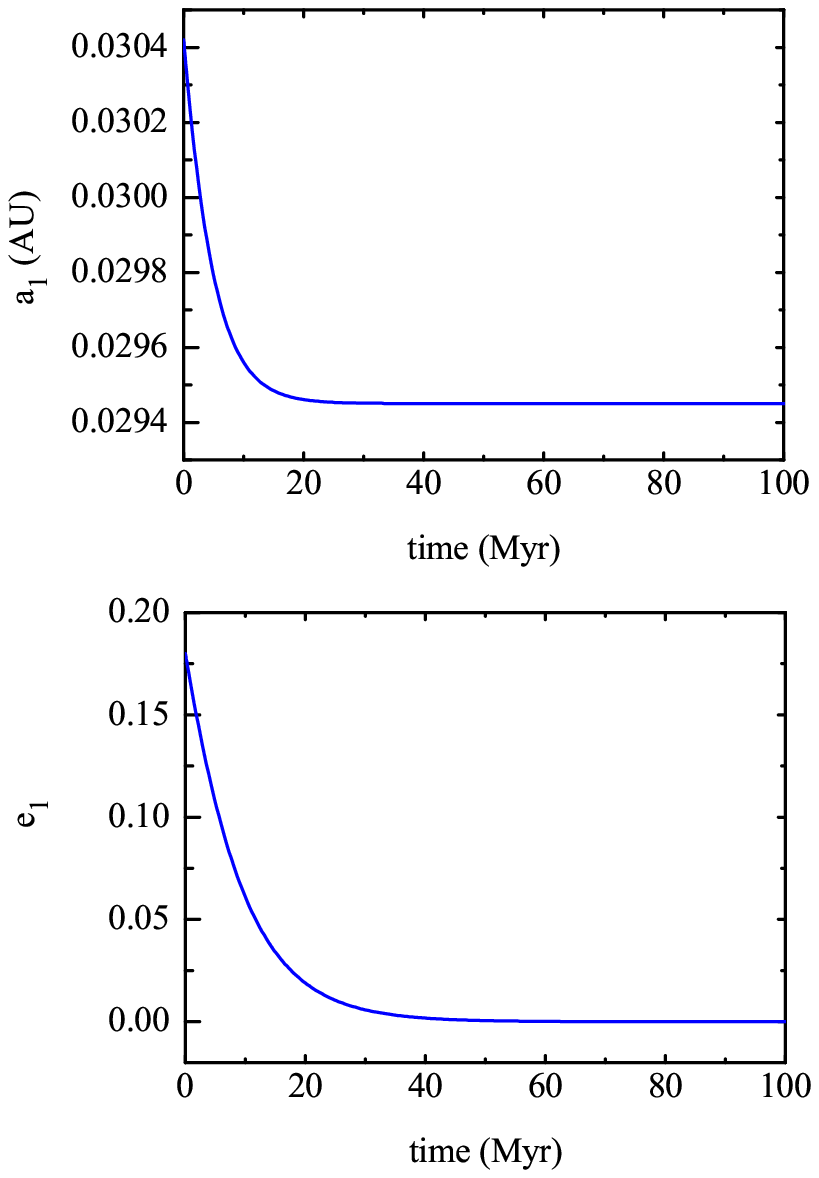,width=3.2in}}\end{center}\vspace*{-8.5mm}
 {\footnotesize {\bf Figure 2}
 \quad Variations of the semi-major axis and eccentricity of WASP-50b in the tidal evolution.}\vspace*{5mm}

A short-period eccentric planet may possess a circular orbit if its
circularization timescale is shorter than the stellar age due to
tidal evolution. The circularization timescale is a vital factor to
reproduce the process of tidal dissipation, which can be described
by (in a single-planet when simply considering the planetary tide
and assuming $a$ as constant) [25]

\begin{equation}
 \tau_{\mathrm{circ}} = -\frac{e}{\dot{e}}  =
 \frac{4}{63}Q_1\Big(\frac{a_1^3}{Gm_0}\Big)^{1/2}\frac{m_1}{m_0}\Big(\frac{a_1}{R_1}\Big)^5
  \end{equation}
For $Q_1^\prime$ = $10^5$ (where $Q_1$ $\simeq$ 26666.7 and $k_1$
$\simeq$ 0.4), the circularization timescale of WASP-50 system is
$\sim$ 6.1 Myr, which is consistent with the numerical simulations
of 9.4 Myr.

The semi-major axis continues to suffer from tidal decay even its
eccentricity is reduced to be about zero owing to the stellar tide
(Figure 3). Subsequently, WASP-50b may disintegrate by the tides
near the Roche limit at $\sim$ 0.01 AU in 0.4, 4 and 40 Gyr when
taking the values of $Q^\prime_0$ as $10^6$, $10^7$ and $10^8$,
respectively. According to the theory, the tidal inspiral time of a
planet is written as [26]
\begin{equation}
 \tau_{a} \approx  \frac{1}{48}\frac{Q^\prime_0}{n_1}\Big(\frac{a_1}{R_0}\Big)^5\frac{m_0}{m_1}
 \label{tao-a}
\end{equation}
which represents the remaining lifetime of the planet. For example,
the very time from the current location to its Roche limit. As
known, $Q^\prime_0$ plays a very important part in tidal evolution,
however, $Q^\prime_0$ is still unknown. Hence, values of
$Q^\prime_0$ in the investigated exo-planetary systems are always
referred to those of solar system. For Jupiter-like system as
WASP-50, when $Q^\prime_0$ = $10^6$, $10^7$, $10^8$, $\tau_{a}$ =
0.29, 2.9, 29 Gyr, which is in good agreement with those numerical
results.

\begin{center}
 \centerline{\psfig{figure=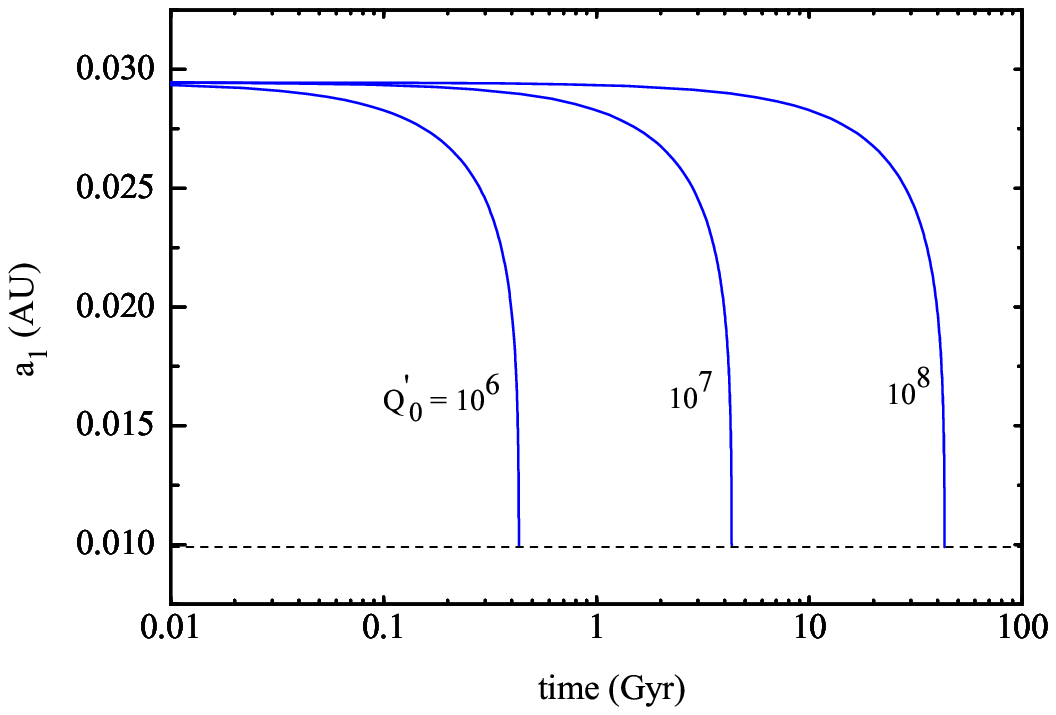,width=3.2in}}\end{center}\vspace*{-8.5mm}
 {\footnotesize {\bf Figure 3}
 \quad The stellar tidal evolution of WASP-50b from the current location with
 various $Q^\prime_0$ (The dash line means the Roche limit at $\sim$ 0.01 AU).}

\subsection{GJ 1214 System}

GJ 1214b is a super-earth detected by transit method in 2009 with a
mass and radius of $6.55~M_{\oplus}$ and $2.678~ R_{\oplus}$
respectively, indicating that its size is between the Earth and ice
giant planets of the solar system. As for planetary structure, GJ
1214b resembles a planet with a composition of water surrounded by a
hydrogen-helium envelope with only 0.05\% of the total planetary
mass [27]. Present-day observations show that GJ 1214b is now
orbiting its host star (0.153 $M_{\odot}$, 0.210 $R_{\odot}$) at
0.014 AU (the smallest semi-major axis in the exo-planetary system
 thus far, see Table 2) with an upper limit of eccentricity (0.27)
[27]. Carter et al. (2011) [28] further constrained the eccentricity
of GJ 1214b as 0.138 on the bases of follow-up light curves.
However, the planetary eccentricity should have been deduced to be
zero in a very close-in orbit due to tidal interaction, however this
inferring is incompatible with a moderate eccentricity of GJ 1214b
as mentioned. One possible explanation is that there are so many
unconfirmed factors during the process of tidal dissipation that
current observations cannot provide a reasonable constraint on GJ
1214b's eccentricity. In this sense, we then extensively investigate
tidal evolution for GJ 1214b by adopting a variety of
eccentricities, in attempt to understand a possible constraint on
its initial eccentricity. Furthermore, we again perform additional
simulations to explore the tidal evolution for this system, using an
alternative $Q^\prime_1$ since the factor is not well determined, to
examine a possible range of $Q^\prime_1$ during the evolution.

\vspace{10mm}\noindent {\small Table 2\quad
Orbital and physical parameters of GJ 1214 system [27]}\\
\vspace{-5mm} {\footnotesize
\begin{center}
\begin{tabular}{lccccc}
\hline
Body     &  Mass                & Radius                 & Semi-major  axis  &  Eccentricity \\
\hline
GJ 1214  & 0.153  $M_{\odot}$   &   0.210 $R_{\odot}$  &$-$  &    $-$  \\
GJ 1214b & 6.55   $M_{\oplus}$  &  2.678 $R_{\oplus}$  &  0.014 AU &  $<$0.27  \\
 \hline
\end{tabular}
\end{center}}\vspace{4mm}

First, to simulate the tidal evolution of various eccentricities of
GJ 1214b, we assume the initial orbital elements as follows:
$a_{1~\mathrm{\mathrm{ini}}}$ = 0.0168 AU, $e_{1~\mathrm{ini}}$ =
0.3, 0.35, 0.4, 0.45. Taking $Q_1^\prime$ = 100 (a typical value for
terrestrial planets), Figure 4 shows variations of the semi-major
axis and eccentricity of GJ 1214b for the variety of the initial
eccentricities, considering both tide and GR raised by the host
star. The outcomes indicate that final orbits are somewhat dependent
on $e_{1~\mathrm{ini}}$, where the smaller the final semi-major axis
(the shorter the circularization timescale), the larger
$e_{1~\mathrm{ini}}$. At the time GJ 1214b finally located at 0.014
AU after experienced tidal decay, the corresponding eccentricities
are deduced to be 0.104, 0.195, 0.267 and 0.330 (marked by red dots
in Figure 4) respectively, of which the former three cases can meet
the upper limit of observed eccentricity (0.27).  Hence, the
possible largest initial eccentricity is 0.4 if the initial
semi-major axis is 0.0168 AU.

\vspace*{-5mm}
\begin{center}
 \centerline{\psfig{figure=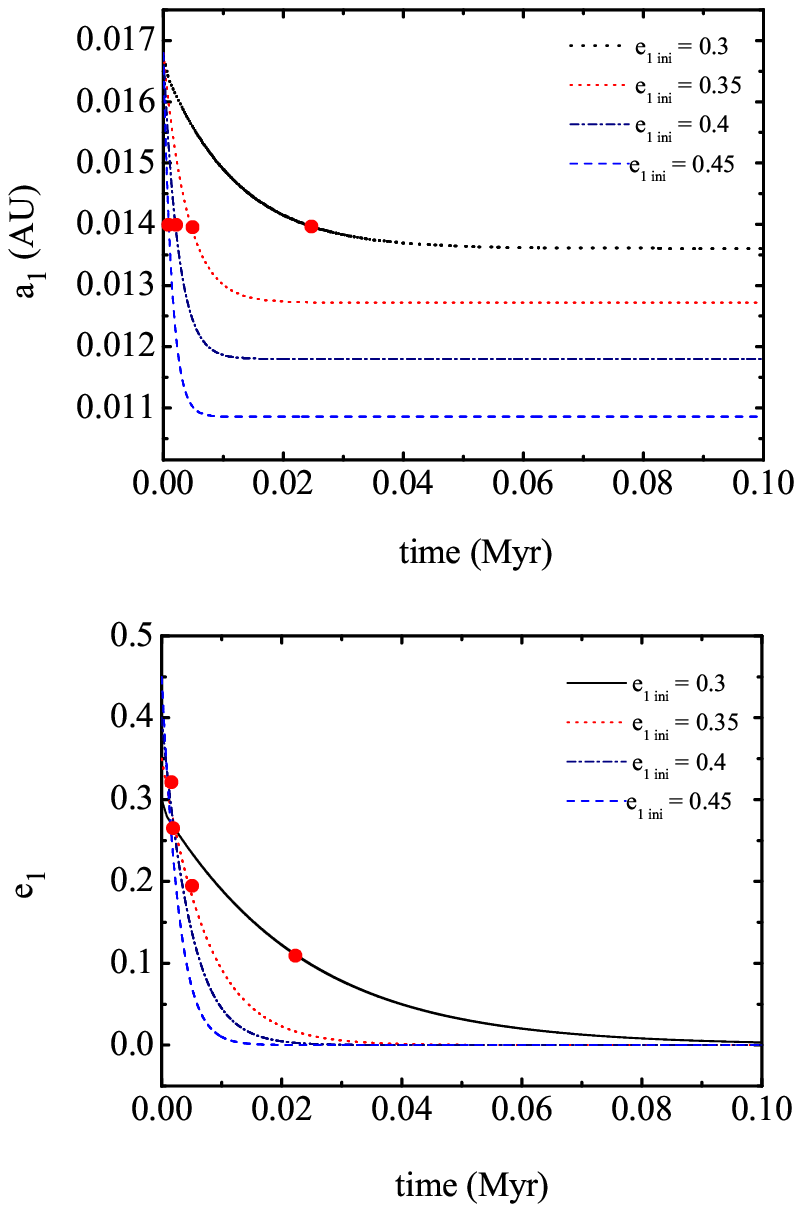,width=3.0in}}\end{center}\vspace*{-8.5mm}
 {\footnotesize {\bf Figure 4}
 \quad Variations of the semi-major axis and eccentricity of GJ
1214b in tidal evolution. The red dots show the current locations
and its corresponding possible eccentricities.}

\begin{center}
 \centerline{\psfig{figure=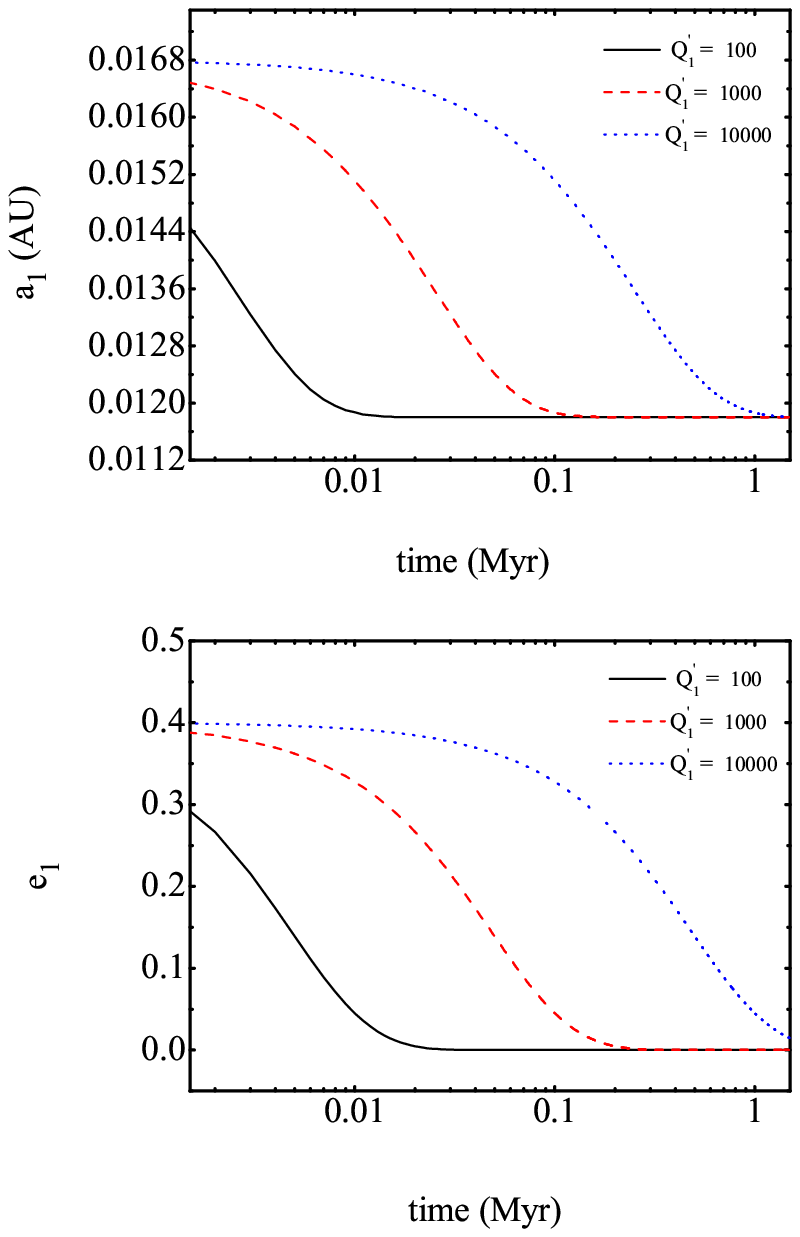,width=3.2in}}  \end{center}\vspace*{-8.5mm}
 {\footnotesize {\bf Figure 5}
 \quad Tidal evolution of GJ 1214b with various $Q^{\prime}_1$.}\vspace*{5mm}

The  circularization timescale is estimated to be $\sim$ 8702.6 yrs
according to equation (10), which is in accord with our numerical
result ($\sim$ $10,000$ yrs) where $Q_1^\prime$ = 100 (with $Q_1$ =
20, $k_1$ = 0.3) in the third case ($a_{1~\mathrm{ini}}$ = 0.0168
AU, $e_{1~\mathrm{ini}}$ = 0.4). Results of this run show that the
final semi-major axis moves about 0.0136 AU on a circular orbit over
$\sim$ 5 $\times$ $10^5$ yrs. If it takes  GJ 1214b so short time to
circularize its orbit, the statistical results may imply that most
exoplanets have circular orbits, but in fact many eccentric planets
have been discovered so far. Therefore, the circularization
timescale of GJ 1214b does not seem to be so short possibly because
the adopted $Q_1^\prime$ = 100 is not reasonable, denoting that the
real value of $Q_1^\prime$ would be likely to be much larger than
100. To make sure a potential range of $Q_1^\prime$, two additional
runs (where $Q^{\prime}_1$ = 1000, 10000) are further investigated
in the study (Figure 5). As seen, $Q^{\prime}_1$ simply plays an
important role in circularization timescale (where a longer
circularization timescale is assocated with a larger
$Q^{\prime}_1$), but has little influence on the present-day
eccentricity and final locations. Substantially speaking, it takes
GJ 1214b about 0.14 Myr and 1.4 Myr to evolve into a circular orbit
from one of possible current cases ($a_1$ = 0.014 AU, $e_1$ = 0.267)
when $Q^{\prime}_1$ = 1000 and 10000, respectively. Therefore, it
seems to be reasonable for GJ 1214b that $Q^{\prime}_1$ is much
larger than 100, indicating that this planet may own an alternate
structure differing from that of a typical terrestrial planet of the
solar system. In this sense, GJ 1214b is likely to be in a process
of undergoing tidal circularization at a slow rate, and its
circularization timescale is proportional to $Q^{\prime}_1$. Hence,
from the simulations, we may infer that a majority of close-in
eccentric exoplanets observed now have larger tidal dissipation
factors than those of the planets in the solar system, leading to a
fact that they may still bear apparent eccentricities far from zero.

  \begin{center}
 \centerline{\psfig{figure=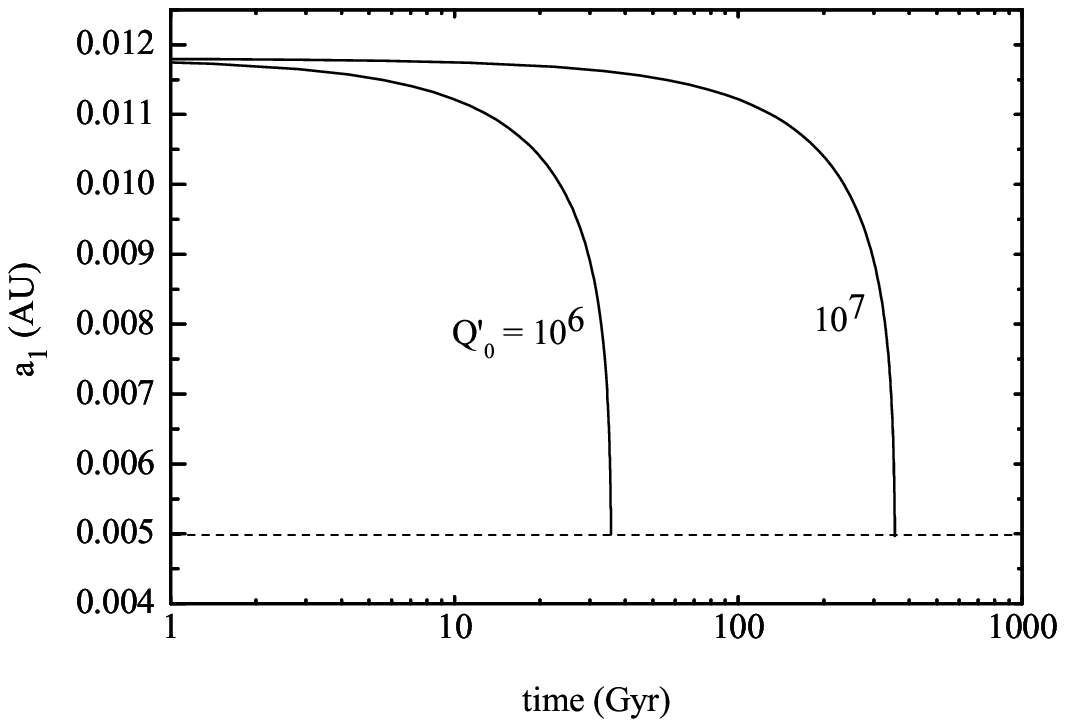,width=3.2in}}\end{center}\vspace*{-8.5mm}
 {\footnotesize {\bf Figure 6}
 \quad The stellar tidal evolution of GJ 1214b with various $Q^\prime_0$.}\vspace*{5mm}

Herein we take $Q^{\prime}_1$ = 100 to estimate the remaining
lifetime of GJ 1214b, in this case its final orbit locates at 0.0118
AU on a circular orbit. Choosing $Q^\prime_0 = 10^6$, $10^7$, the
calculated timescale of remaining lifetime is about 28.8 and 288.0
Gyr respectively, according to equation (11), and this is consistent
with numerical results (Figure 6). The remaining lifetime of GJ
1214b is longer than that of WASP-50b because of a relatively lower
mass of GJ 1214. If the main-sequence lifetimes of stars such as the
type of GJ 1214, with a mass of only 0.15 $M_{\odot}$, are 100 Gyr
by taking $Q^\prime_0$ = $10^6$, GJ 1214b will eventually be affected
by stellar tide, however, it will survive throughout the stellar
lifetime for $Q^\prime_0$ = $10^7$.

\subsection{CoRoT-7 System}

CoRoT-7b is the first super-earth discovered with a determined
radius and mass (8.0~$M_\oplus$, 1.58~$R_\oplus$) about its host
star at a close-in circular orbit (0.0172~ AU), companied with a
second Earth-like planet CoRoT-7c (13.6 $M_\oplus$, 2.39 $R_\oplus$)
at 0.046 AU (Table 3) [29, 30]. Hence, the CoRoT-7 system, hosting
two terrestrial planets, motivates a great many of researchers to
explore its formation and dynamical evolution.

\vspace{4mm}\noindent {\small Table 3\quad Orbital and physical data
of CoRoT-7 system [29-31]}\\\vspace{-4mm} {\footnotesize
\begin{center}
\begin{tabular}{lccccc}
\hline
Body     &  Mass           & Radius          & Semi-major axis  &  Eccentricity\\
\hline
 CoRoT-7      &   0.93 $M_{\odot}$   &  0.87 $R_{\odot}$   &    $-$ \\
 CoRoT-7b     &   8.0  $M_{\oplus}$  &  1.58 $R_{\oplus}$  & 0.017 AU  &  0 \\
 CoRoT-7c     &   13.6 $M_{\oplus}$  &  2.39 $R_{\oplus}$  & 0.046 AU  &  0 \\
 \hline
\end{tabular}
\end{center}}\vspace{4mm}

\vspace{-4mm}
\begin{center}
 \centerline{\psfig{figure=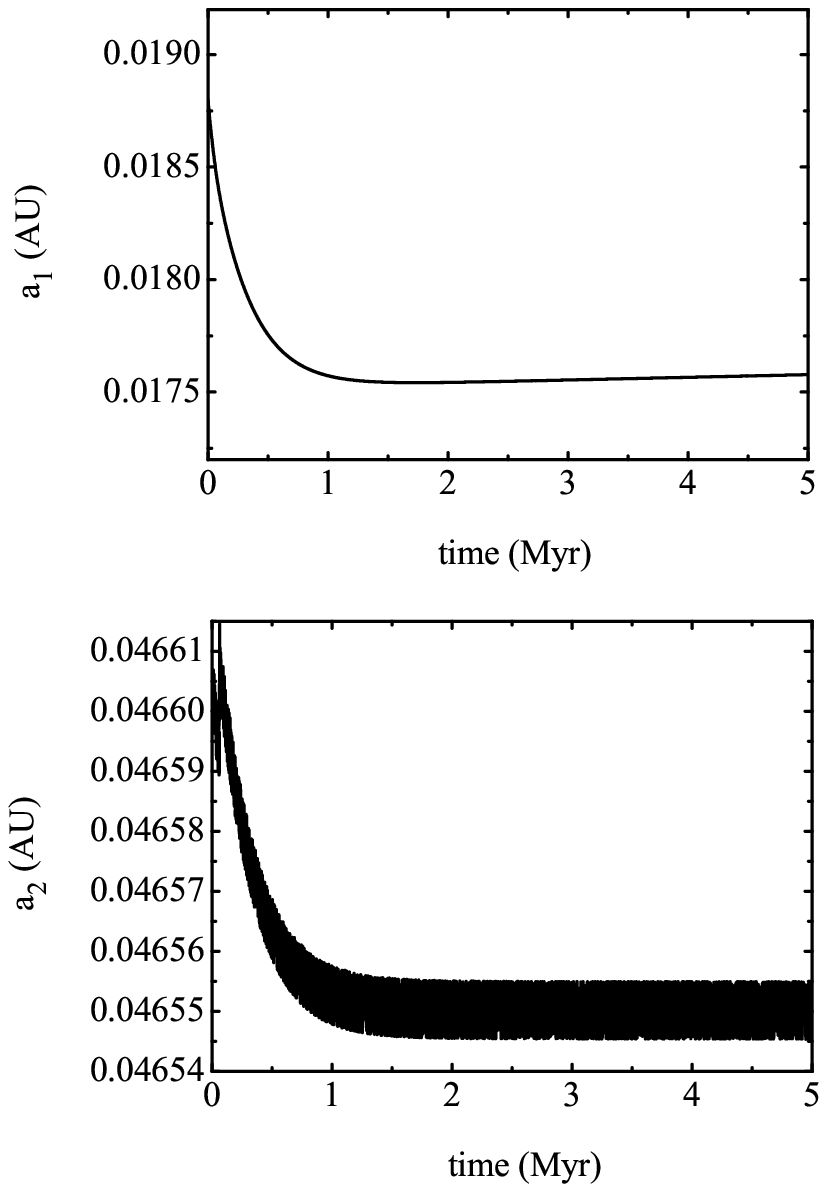,width=3.2in}}\end{center} \vspace*{-8.5mm}
 {\footnotesize {\bf Figure 7}
 \quad Variations of semi-major axes of CoRoT-7 system in tidal evolution.}

\begin{center}
 \centerline{\psfig{figure=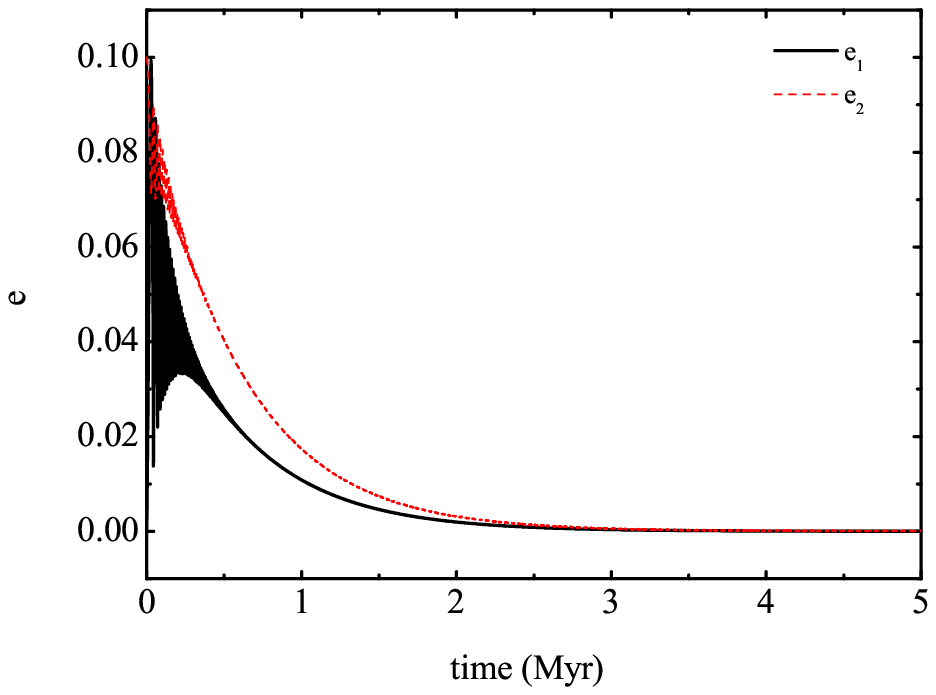,width=3.2in}}\end{center}\vspace*{-8.5mm}
{\footnotesize {\bf Figure 8}
 \quad The variations of eccentricities of CoRoT-7 system in tidal evolution.}\vspace*{5mm}

Considering the total angular momentum as invariable, the initial
orbital elements are assumed as $e_{1~\mathrm{ini}}$ = 0, $e_{2~
\mathrm{ini}}$ = 0.1, $a_{1~\mathrm{\mathrm{ini}}}$ = 0.0188 AU,
$a_{2~\mathrm {ini}}$ = 0.0466 AU. Figures 7 and 8 show the tidal
evolution of two planets under consideration of  both tide and GR
($Q^\prime_1$ = $Q^\prime_2$ = 100). The semi-major axes and
eccentricities of two planets are both decreasing due to tidal
effects. The initial eccentricity of CoRoT-7b is pumped up to a
moderate value of 0.1 due to the excitation of the outer eccentric
companion, then it is relaxed within a short time of $\sim$ 0.5 Myr,
next approaches a quasi-equilibrium state. The excitation of
eccentricity is essential for the inner planet to undergo tidal
decay and circularization as the tidal dissipation requires a
nonzero eccentricity according to equation (1)-(3). After the
relaxation progress, the orbit of the inner planet becomes smooth
and continues to suffer from tidal decay and circularization. As a
result that it stops near the current location at 0.0175 AU in a
circular orbit. Concurrently, the outer planet of CoRoT-7c still
migrates towards the star at a much slower rate comparing to that of
the inner planet owning to the farther distance from its host star.
Although both of the system run towards the star at various
migration rate, resulting in a much wider separation between them, a
pair of circular orbits have finally formed at about 2 Myr.

 \vspace*{-3mm}
\begin{center}
 \centerline{\psfig{figure=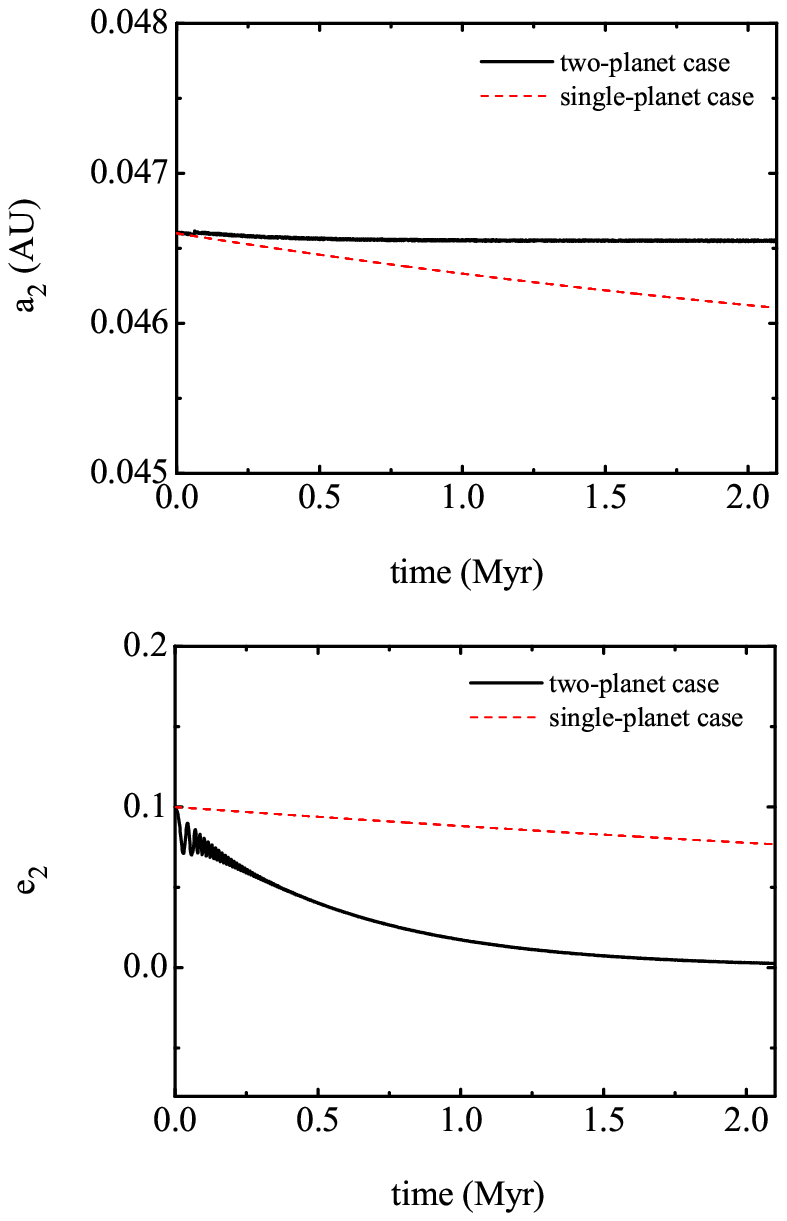,width=3.2in}}\end{center}
 \vspace*{3mm}
 {\footnotesize {\bf Figure 9}
 \quad Tidal evolution of CoRoT-7c in single-planet and two-planet cases.}\vspace*{3mm}

In addition, we carry out additional runs to investigate the tidal
evolution of only a single outer planet to understand the outer
planet's orbital decay and circularization. In this case, orbital
shrinkage is much quicker but circularization is much slower than
that of the two-planet case (Figure 9). As a consequence, the
decreasing of $e_2$ is mainly produced by the tidal decay and
circulation of the inner planet, as a result of coupled tidal and
gravitational interactions rather the contributions from its own
tidal dissipation. However, to conserve the total angular momentum
in the planet-star system, the tidal decay of the outer planet is
damped in presence of the inner planet. In theory, Mardling (2007)
[32] studied the secular evolution of the outer planet in a
two-planet system, expressing the variation of $e_2$ as

\begin{equation}
\dot{e}_2=-\frac{\lambda}{\tau_{\textrm{\scriptsize
circ}}}\,\frac{e_2}{F(e_2)},
\end{equation}
where $\tau_{\textrm{\scriptsize circ}}$ = $e_1/\dot{e}_1$ is the
circularization timescale of the inner planet in a single-planet
system, and $\dot{e}_1$ is given by equation (12),
$\lambda\equiv(25/16)(m_1/m_2)(a_1/a_2)^{5/2}$,
  and $F(e_2)\equiv\varepsilon_2^3(1-\alpha\varepsilon_2^{-1}+\gamma\varepsilon_2^3)^2$,
  $\alpha\equiv(m_1/m_2)(a_1/a_2)^{1/2}$.

\begin{center}
 \centerline{\psfig{figure=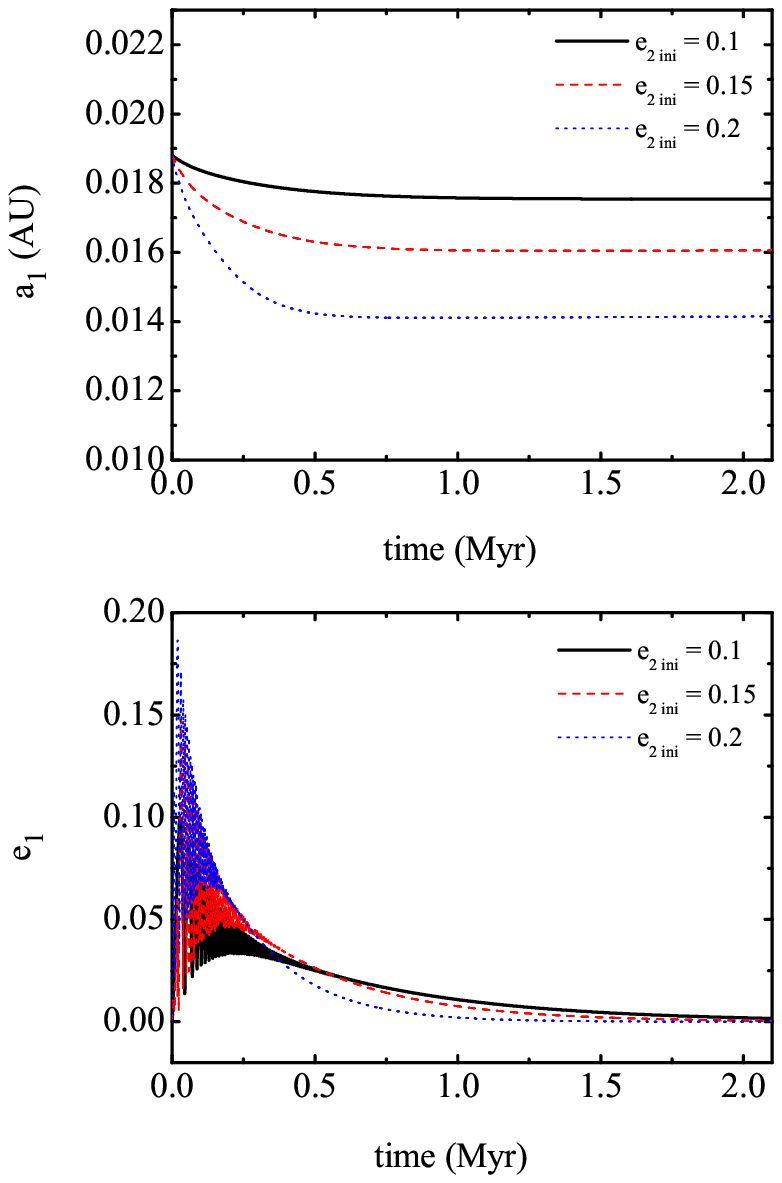,width=3.2in}}\end{center}\vspace*{-8.5mm}
 {\footnotesize {\bf Figure 10}
 \quad Tidal evolution of CoRoT-7b with a variety of $e_{2~\mathrm{ini}}$.}\vspace*{3mm}

Qualitatively, our results agree with those of Rodr\'{\i}guez et al.
(2011) [16], but the resulting orbits we obtained differ from theirs
because of the adoption of the initial orbital elements, as the
initial parameters have an important part in its final orbit. In
this sense, three runs of simply changing $e_{2~\mathrm{ini}}$ =
0.1, 0.15, 0.2, respectively, are again performed to explore the
situation of orbital evolution for a two-planet system. Figures 10
and 11 show that both semi-major axes and eccentricities are
influenced by $e_{2~\mathrm{ini}}$, as a result, the final locations
of the inner and the outer planets are at 0.0175, 0.0166, 0.0140 AU
and 0.04656, 0.04651, 0.04647 AU, respectively, and the changes in
final semi-major axis of the outer planet is smaller as it is far
away from the star. The final semi-major axes are smaller with
larger $e_{2~\mathrm{ini}}$, and the changes of their eccentricities
can be ignored since all of them are deduced to be zero within
$\sim$ 2 Myr, implying that their final locations are mainly
dominated by $e_{2~\mathrm{ini}}$. Furthermore it provides concrete
evidence that tidal effects and mutual gravitational interactions
are coupled with each other during the secular evolution.

\vspace*{-5mm}
\begin{center}
 \centerline{\psfig{figure=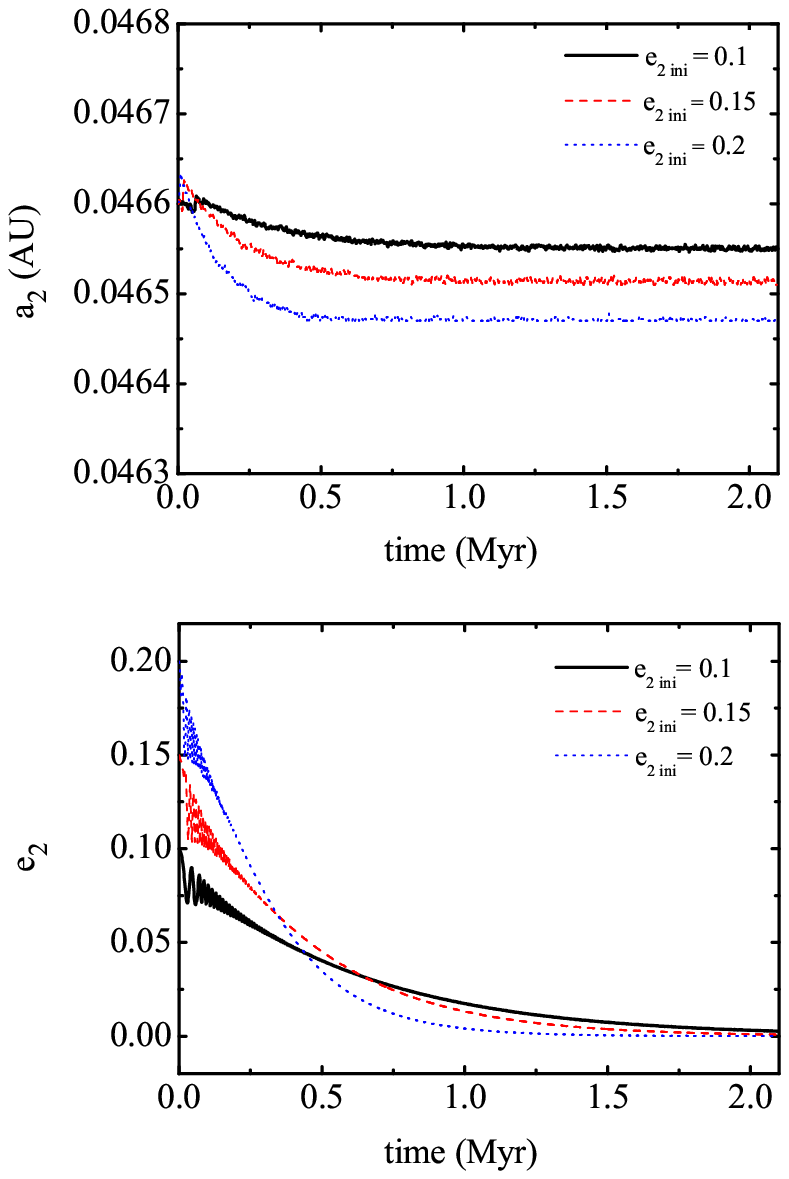,width=3.2in}}\end{center} \vspace*{-8.5mm}
 {\footnotesize {\bf Figure 11}
 \quad Tidal evolution of CoRoT-7c with a variety of $e_{2~\mathrm{ini}}$.}\vspace*{3mm}

Figure 12 shows the stellar tidal evolution from the current orbit
for $Q_0^\prime$ = $10^5$, $10^6$, $10^7$, respectively. For
example, if taking $Q_0^\prime$ = $10^6$, CoRoT-7b will be
demolished in less than 1 Gyr, which is in concordance with the
evaluated tidal inspiral timescale $\tau_{a_1}$ = 0.4 Gyr. However,
CoRoT-7c will still remain survival in the stellar lifetime due to
its extremely far proximity from the host star by taking the above
values of $Q_0^\prime$.

\vspace*{-5mm}
 \begin{center}
 \centerline{\psfig{figure=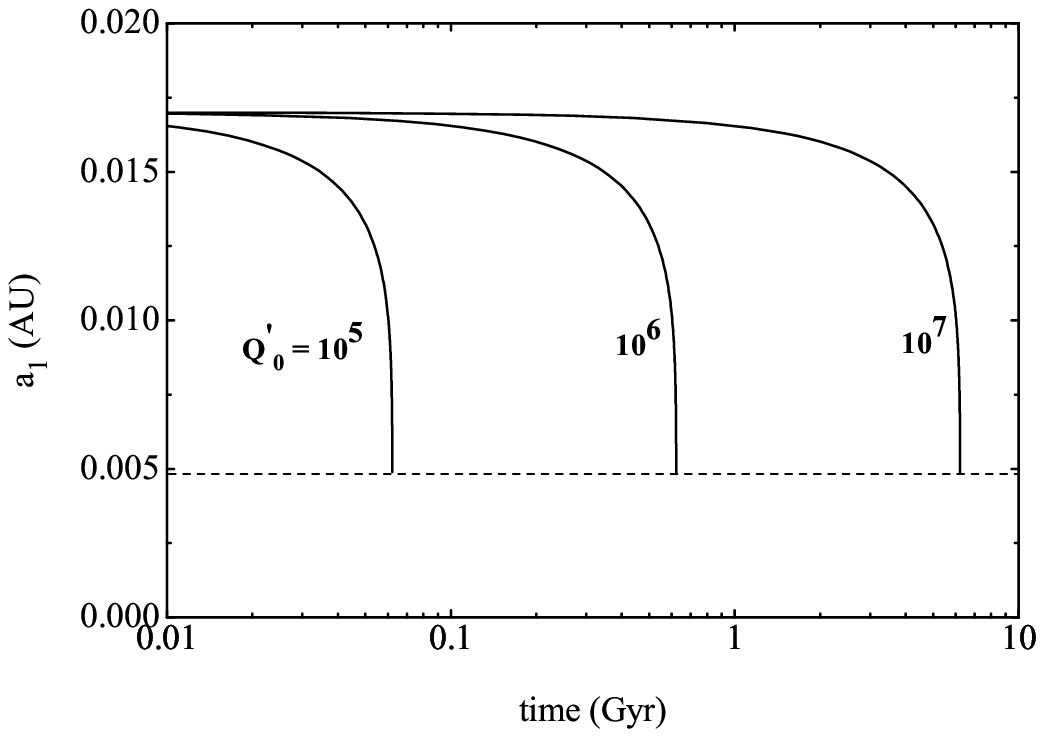,width=3.2in}}\end{center}\vspace*{-8.5mm}
 {\footnotesize {\bf Figure 12}\quad The stellar tidal evolution of
CoRoT-7b with various $Q^\prime_0$.}

\subsection{Summary and Discussion}

In this work, we numerically simulate two single-planet systems
including one hot Jupiter-like planetary system (WASP-50) and hot
super-earth system (GJ 1214), and a two-planet Earth-like system
(CoRoT-7). We may summarize the main results below:

WASP-50b may arrive at its final orbit with a zero eccentricity in
about 50 Myr by suffering from tidal decay and circularization. For
GJ 1214 system, alternative initial eccentricities in simulations
are investigated to make clear the tidal evolution for the
super-earth GJ 1214b. As a result that the final place is more
closer to the star and the eccentricity of the current location is
much larger with a larger initial eccentricity in tidal process, and
an upper limit of initial eccentricity (0.4) is given at the initial
location (0.0168 AU). In addition, $Q^\prime_1$ may have a direct
influence on the circularization timescale of GJ 1214b, implying
that $Q^\prime_1$ of GJ 1214b could be probably much larger than
that of terrestrial planets (100), so as to explain why GJ 1214b
still bears an apparent eccentricity in such a close-in orbit.
For two-planet system CoRoT-7, tidal evolution induces two planets
to approximatively reach their current locations with zero
eccentricities respectively, although both of them suffer from
mutual influences due to coupled affects of tidal and gravitational
interaction, which can be explained by the conservation of angular
momentum of the planet-star system. When the eccentricity goes down
to zero, the stellar tide then begins to work and causes the orbital
decay of the planet until it enters the region of Roche limit.
Considering $Q^\prime_1$ = $10^6$, the remaining lifetimes of
WASP-50b, GJ 1214b and CoRoT-7b are about a few billion years in
comparison with the stellar ages, however, CoRoT-7c may survive
steadily owing to relatively farther distance from its host star.

In this paper, GR is not analyzed although it is numerically
considered in tidal evolution, because GR has no effect on secular
variations of semi-major axis and eccentricity [33]. Herein our
numerical simulations are based on several assumptions:

Firstly, tidal dissipation factor $Q^\prime_1$ (or $Q^\prime_0$) is
adopted from an equilibrium tidal theory, which is independent of
tidal frequency and amplitude; secondly, $Q^\prime_1$ is referred to
those values of the solar system. For instance, taking $10^{5}$ and $100$
for Jupiter-like and terrestrial planets, respectively; thirdly, the
stellar tide arising from the planet exerting on the host star is
ignored during tidal decay and circularization;

Finally, the synchronization is assumed since the spin period of the
planet is unknown and generally the synchronization timescale is
much shorter than a secular tidal evolution. Thus, all assumptions
may produce a deviation from an authentic dynamical evolution for
the investigated systems. Circularization timescales of some
planetary systems may be unreasonable when compared to the ages of
planets due to the improper $Q^\prime_1$, as well as the planetary
ages are inconsistent with their stellar ages caused by the
uncertain $Q^\prime_0$. Therefore, new constraints of $Q^\prime_1$
and $Q^\prime_0$ are expected from observations so as to understand
tidal dissipation well. In addition, the future new models of
planetary formation and migration are of assistance to restrain the
starting setup of the reasonable initial configurations adopted in
numerical simulations that the final orbits depend on. Hence, until
all above issues are well resolved, one complete scenario from
planet formation, migration and tidal evolution to generation of the
final planetary configuration will help us to understand how hot
Jupiter-like and Earth-like planets are formed.

\Acknowledgements{\bahao We thank Professor ZHOU Jilin at Nanjing
University for valuable comments and suggestions. This work was
supported by the National Natural Science Foundation of China (Grant
No.10973044, 10833001), the Natural Science Foundation of Jiangsu
Province (Grant No. BK20093411), and the Foundation of Minor Planets
of the Purple Mountain Observatory.}


\normalsize \vskip0.3in\parskip=0mm \baselineskip 18pt
\renewcommand{\baselinestretch}{1.1}\footnotesize\parindent=4mm\bahao


\REF{1\ }http://exoplanet.eu

\REF{2\ }Murray C D, Dermott S F. Solar System Dynamics (New
               York: Cambridge Univ. Press). 1999

\REF{3\ }Jackson B, Barnes R, Greenberg R. Observation evidence for
             tidal destruction of exoplanets. Astrophys J, 2009, 698: 1357--1366

\REF{4\ }Lanza A F. Hot Jupiters and the evolution of stellar
               angular momentum. Astron Astrophys, 2010, 512: 77--91

\REF{5\ }Lissauer J J. Planet formation. Ann Rev Astron Astrophys,
1993, 31: 129--174

\REF{6\ }Rasio F A, Ford E B. Dynamical instabilities and the
            formation of extrasolar planetary systems. Science, 1996, 274: 954--956

\REF{7\ }Zhou J L, Aarseth S J, Lin D N C, et al. Origin and
  ubiquity of short-period earth-like planets: evidence for the
  sequential accretion theory of planet of formation. Astrophys J, 2005, 631: L85¨C-L88

\REF{8\ }Ji J H, Jin S, Tinney C G. Forming close-in earth-like
           planets via a collision-mergermechanism in late-stage planet
           formation. Astrophys J, 2011, 727: L5--L8

\REF{9\ }Jin S, Ji J H. Terrestrial planet formation in inclined
systems: application to the OGLE-2006-BLG-109L
             system. Mon Not Roy Astron Soc, 2011, 418: 1335--1345

\REF{10\ }Lin D, Bodenheimer P, Richardson D. Orbital migration of
           the planetary companion of 51 Pegasi to its present
           location. Nature, 1996, 380: 606--607

\REF{11\ }Goldreich P. Disk-satellite interactions. Astrophys J,
              1980, 241: 425--441

\REF{12\ }Fabrycky D, Tremaine S. Shrinking Binary and Planetary
                  Orbits by Kozai Cycles with Tidal Friction. Astrophys J, 2007, 669: 1298--1315

\REF{13\ }Ford E B, Rasio F A. on the relation between hot jupiters
                and the roche limit. Astrophys J, 2006, 638: L45--L48

\REF{14\ }Rasio F A, Tout C A, Lubow S H, et al. Tidal decay of
                     close planetary orbits. Astrophys J, 1996, 470: 1187-1191

\REF{15\ }Zhou J L, Lin D N C. Migration and Final Location of Hot
                       Super Earths in the Presence of Gas Giants, in IAU Symp. 249,
                        Exoplanets: Detection, Formation and Dynamics, ed. Y.-S. Sun, S.
                      Ferraz-Mello \& J.-L. Zhou (China:Suzhou), 2008, 319: 285-289

\REF{16\ }Rodr\'{\i}guez A, Ferraz-Mello S, Michtchenko T A, et al.
Tidal decay and orbital circularization in close-in two-planet
                  systems. Mon Not Roy Astron Soc, 2011, 415: 2349--2358

\REF{17\ }Dobbs-Dixon I, Lin D N C, Mardling R A. Spin-Orbit
               Evolution of Short-Period Planets. Astrophys J, 2004, 610: 464--476

\REF{18\ }Beutler G. Methods of Celestial Mechanics. Vol.
                I:Physical, mathematical, and numerical principles (Springer,
                Berlin), 2005

\REF{19\ }Mignard F. Moon Planets. 1979, 20: 301

\REF{20\ }Mardling R A, Lin D N C. Calculating the Tidal, Spin,
                    and Dynamical Evolution of Extrasolar Planetary Systems.  Astrophys J,
                    2002, 573: 829--844

\REF{21\ }Chambers J E. A hybrid symplectic integrator that permits
                       close encounters between massive bodies. Mon Not Roy Astron
                       Soc, 1999, 304: 793--799

\REF{22\ }Stoer J, Bulirsch R. Introduction to Numerical Analysis (New York: Springer Verlag), 1980

\REF{23\ }Chambers J E. A hybrid symplectic integrator that permits
close encounters between massive bodies. Mon Not Roy Astron
                       Soc, 1999, 304:793--799

 \REF{24\ }Gillon M, Doyle A P, Lendl M, et al. WASP-50b: a hot
             Jupiter transiting a moderately active solar-type star. Astron Astrophys, 2011,
                  533: 88--95
 \REF{25\ }Goldreich P, Soter S. $Q$ in the Solar System. Icarus, 1966, 5: 375--389

\REF{26\ }Levrard B., Winisdoerffer C., Chabrier G., Falling
                    transiting extrasolar giant planets. Astrophys J, 2009, 692: L9--L13

\REF{27\ }Charbonneau D, Berta Z K, Irwin J. A super-Earth
              transiting a nearby low-mass star. Nature, 2009, 462: 891--894

\REF{28\ }Carter J A, Winn J N, Holman M J, et al. The transit light
                curve project, XIII. sixteen transits of he super-earth GJ 1214b.
                  Astrophys J, 2011, 730: 82--91

\REF{29\ }L\'{e}ger A, Rouan D, Schneider J, et al. Transiting
                  exoplanets from the CoRoT space mission VIII. CoRoT-7b: the first
                       super-Earth with measured radius. Astron Astrophys, 2009, 506: 287-302

\REF{30\ }Queloz D, Bouchy F, Moutou C, et al. The CoRoT-7
                        planetary system: two orbiting super-Earths. Astron Astrophys, 2009, 506: 303-319

\REF{31\ }Ferraz-Mello S, Tadeu D S M, Beaug\'{e} C, et al. On the
                        mass determination of super-Earths orbiting
                        active stars: the CoRoT-7 system. Astron Astrophys, 2011, 531: 161-171

\REF{32\ }Mardling R A. Long-term tidal evolution of short-period
                  planets with companions. Mon Not Roy Astron Soc, 2007, 382: 1768--1790

\REF{33\ }Huang C., Liu L. Analytical solutions to the four
                   post-Newtonian effects in a near-earth satellite orbit. Celest Mech Dyn Astr, 1992, 53: 293--307
\end{multicols}

\end{document}